\documentclass[prl,twocolumn,showpacs]{revtex4}
\usepackage[dvips]{graphicx}
\usepackage{amsmath}

\begin{document}

\title{Coherent Collisions between Bose-Einstein Condensates}

\begin{abstract}
We study the non-degenerate parametric amplifier for matter waves,
implemented by colliding two Bose-Einstein condensates. The coherence of the
amplified waves is shown by observing high contrast interference with a
reference wave and by reversing the amplification process. Since our
experiments also place limits on all known sources of decoherence, we infer
that relative number squeezing is most likely present between the amplified
modes. Finally, we suggest that reversal of the amplification process may be
used to detect relative number squeezing without requiring single-particle
detection.

%
\end{abstract}

\author{J. M. Vogels}
\author{J. K. Chin}
\author{W. Ketterle}
\homepage[Group website: ]{http://cua.mit.edu/ketterle_group/}
\affiliation{Department of Physics, MIT-Harvard Center for Ultracold
Atoms, and Research Laboratory of Electronics,
Massachusetts Institute of Technology, Cambridge, MA 02139}
\maketitle%
%
\catcode`\ä = \active \catcode`\ö = \active \catcode`\ü = \active
\catcode`\Ä = \active \catcode`\Ö = \active \catcode`\Ü = \active
\catcode`\ß = \active \catcode`\é = \active \catcode`\è = \active
\catcode`\ë = \active \catcode`\ô = \active \catcode`\ê = \active
\catcode`\ø = \active \catcode`\ò = \active \catcode`\í = \active
\defä{\"a} \defö{\"o} \defü{\"u} \defÄ{\"A} \defÖ{\"O} \defÜ{\"U} \defß{\ss} \defé{\'{e}}
\defè{\`{e}} \defë{\"{e}} \defô{\^{o}} \defê{\^{e}} \defø{\o} \defò{\`{o}} \defí{\'{i}}%
%
Quantum entanglement has always attracted great interest as it is one of the
key differences between quantum and classical physics. Since it can be
exploited for various novel applications like quantum computation and
precision measurements, there has been a continuing effort to engineer ever
more robust quantum correlated states. Recently, four-wave mixing using
Bose-Einstein condensates (BEC's) has been proposed as a method for creating
two correlated matter waves \cite{deng99}. In that experiment, two source
waves and a third seed wave were used to produce a fourth wave in the
conjugate momentum state, resulting in an amplified seed-conjugate wave
pair. For large amplification, the relative number fluctuations between the
amplified waves can become significantly reduced, and this was recently
achieved \cite{Voge2002}. In optics, this is known as non-degenerate
parametric amplification \cite{note1}, where it is also used for generating
similarly squeezed photon beams. Once created, such two-mode number squeezed
states could be used for Heisenberg-limited interferometry where phase
shifts up to $1/N$ accuracy could be measured \cite{holl93, bouy97}. As the
occupation number $N$ of the states increases, the advantage gained over the
standard shot noise limit of $1/\sqrt{N}$ becomes quite significant.

Entanglement between the generated matter waves is possible only if the
collisional four-wave mixing process is coherent. While the coherence of
collisions has been observed before in mean field effects \cite{Grein2002,
Dalf1999}, it has not been demonstrated for the more common case of momentum
changing elastic collisions. By itself, four wave mixing does not require or
guarantee coherence, since it will still occur for condensates with short
coherence lengths \cite{Dett2001}. Moreover, even though previous four wave
mixing experiments \cite{deng99, Voge2002} utilized two coherent
condensates, they did not explicitly show that coherence was maintained
throughout the collisional process.

In this paper, we demonstrate the coherence of elastic collisions by
observing high contrast interference between the conjugate wave and a
reference wave. Due to the strong coherence retained, we also observe a
reversal of the amplification process. The success of these two experiments
places an upper bound on the decoherence effects in our experiments,
providing evidence that a significant amount of relative number squeezing is
present in the pair-correlated beams. The reversibility of the amplification
process further suggests a possible way of directly detecting the two-mode
number squeezing and entanglement between the two correlated states, without
requiring sub-shotnoise detection.

In the Heisenberg picture, the evolution of the annihilation operators $%
a_{3} $, $a_{4}$ of a mode pair that is parametrically amplified by the
source modes $a_{1}$, $a_{2}$ is given by

\begin{eqnarray}
\frac{da_{3}}{dt} &=&-\frac{8\pi \hbar a}{Vm}ia_{4}^{\dagger }a_{1}a_{2}-i%
\frac{1}{2\hbar }\Delta \varepsilon \text{ }a_{3}  \label{4wmeq} \\
\frac{da_{4}}{dt} &=&-\frac{8\pi \hbar a}{Vm}ia_{3}^{\dagger }a_{1}a_{2}-i%
\frac{1}{2\hbar }\Delta \varepsilon \text{ }a_{4}  \notag
\end{eqnarray}
where $V$ is the volume, $m$ is the mass, $a$ is the s-wave scattering
length and $\Delta \varepsilon $ is the energy mismatch between the source
modes and the amplified modes \cite{Voge2002}. The solution to these
equations for highly occupied source modes is an exponential growth with a
rate constant $\eta =\sqrt{(\frac{2\overline{\mu }}{\hbar })^{2}-(\frac{%
\Delta \varepsilon }{2\hbar })^{2}}$ where $\overline{\mu }$ is the
geometric mean of the chemical potentials of the two source waves. In our
experiment, the multiple waves used were generated from a single BEC using
two-photon stimulated optical transitions \cite{kozu1999}. The evolution of
annihilation operators $a,$ $b$ for modes which are coupled by such a Bragg
transition is given by

\begin{eqnarray}
\frac{da}{dt} &=& -\Omega ib  \label{braggeq} \\
\frac{db}{dt} &=& -\Omega ^{\ast }ia  \notag
\end{eqnarray}
where $\Omega $ is the coupling strength. In both the Bragg and
amplification process, the generated wave incurs an overall phase shift of 90%
$^{\circ }$ relative to its source, a detail which becomes important for the
interference experiment.

\begin{figure}[tbp]
\includegraphics[bb=3in 3in 5.6in 8in,width=30mm]{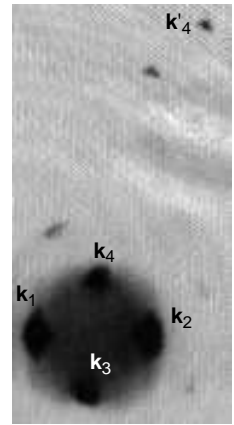}
\caption{The two source waves $\mathbf{k}_{1}$, $\mathbf{k}_{2}$, the seed $%
\mathbf{k}_{3}$, the conjugate wave $\mathbf{k}_{4}$, and the detection
state $\mathbf{k}_{4}^{\prime }$ after 20 ms ballistic expansion. Also seen
is the collisional halo around all four waves as well as several small
unlabelled waves, generated by off-resonant scattering. The field of view is
1.0 x 1.9 mm.}
\label{wavesfig}
\end{figure}

The experimental setup was almost identical to the one used in \cite
{Voge2002}. Sodium condensates of about 13 million atoms each were created
in a cylindrical magnetic trap with an axial frequency of 20 Hz and a radial
frequency of 40 Hz. They had a Thomas-Fermi radius of 30 $\mu $m radially
and 60 $\mu $m axially, a chemical potential of $h$ $\times $ 1.8 kHz and a
speed of sound of 6 mm/s. From an initial condensate \textbf{k$_{1}$}, a
small seed \textbf{k$_{3}$} with a velocity of 14 mm/s was generated using a
20 $\mu $s Bragg (seed) pulse (see Fig.~\ref{wavesfig}). In the next 40 $\mu 
$s, a second source wave \textbf{k$_{2}$} with a velocity of 20 mm/s was
split off from \textbf{k$_{2}$} using a $\pi /2$ pulse. All Bragg beams in
our experiment were detuned from the sodium D$_{2}$ line by 40 GHz to avoid
spontaneous Rayleigh scattering. Subsequent collisions between \textbf{k$%
_{1} $} and \textbf{k$_{2}$} acted as a parametric amplifier for the seed
wave \textbf{k$_{3}$}, creating a fourth conjugate wave \textbf{k$_{4}$} in
the momentum conserving state. Both \textbf{k$_{3}$} and \textbf{k$_{4}$}
modes then grew exponentially with an expected rate of 17~ms$^{-1}$ on
average accross the condensate. At the same time, the parametric amplifier
also enhanced spontaneous scattering into initially unseeded modes, giving
rise to the collisional halo seen in Fig.~\ref{wavesfig} \cite{chik00}. The
resulting loss of atoms from the source waves limited the four wave mixing
process to 400 $\mu $s.

After 80 $\mu $s of amplification, we probed the phase coherence of the
conjugate wave \textbf{k$_{4}$} by interfering it with a reference wave. For
optimal interference, the reference wave needed to have a well defined
relative phase and be identical in size and momentum to \textbf{k$_{4}$}.
The reference wave was outcoupled from \textbf{k$_{2}$} using another 20 $%
\mu $s Bragg (seed) pulse. By changing the phase of one of the Bragg beams,
the phase of the reference wave relative to \textbf{k$_{4}$} was varied. The
resulting occupation of the conjugate momentum state \textbf{k$_{4}$} was
monitored by selectively transferring a fixed fraction of atoms in \textbf{k$%
_4$} to a detection state \textbf{k$_{4}^{\prime }$} (see Fig.~\ref{wavesfig}%
). For this, a 40 $\mu $s Bragg pulse with large momentum transfer (2 $\hbar
k$) was used. This ensured that the \textbf{k$_4^{\prime}$} atoms were
well-separated during ballistic expansion and their number was easily
determined. Direct number measurement of the mode \textbf{k$_{4}$} was not
possible since the amplification process continued during ballistic
expansion.

\begin{figure}[tbp]
\includegraphics[bb=0.5in 3.5in 7.5in 9.5in,width=60mm]{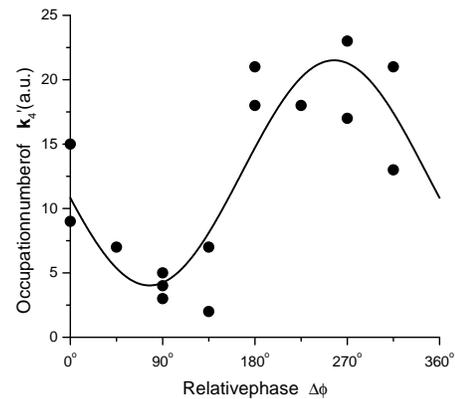}
\caption{Coherence of the conjugate wave. Shown is the interference of the
conjugate wave with a reference wave. The relative phase shift was
introduced by shifting the phase of the rf-wave which drives the
lower-frequency acousto-optic modulator in the Bragg setup. }
\label{fringefig}
\end{figure}

The interference fringe obtained is plotted in Fig.~\ref{fringefig}. By
fitting a sinusoidal curve to the raw data, we obtained a fringe contrast of
68 $\pm $ 11 \%. This proves that the conjugate wave maintains the phase
relation established during amplification and is in a single quantum state.
In addition, we expected destructive interference to take place at 90$%
^{\circ }$ due to the accumulation of all the 90$^{\circ }$ phase shifts
introduced by the four wave mixing and Bragg processes (see eqs.\ \ref{4wmeq}%
, \ref{braggeq}). Our value of 77 $\pm $ 9$^{\circ }$ agrees fairly well
with this, and the small discrepancy could be due to the Bragg beams not
being fully resonant. It is also interesting to note that the position of
the fringe minimum is a direct signature of the positive sign of the
scattering length $a$, since a negative scattering length would have
resulted in a minimum at 270$^{\circ }.$ Usually, only the cross section,
which is proportional to $a^{2}$, is measured using these momentum changing
collisions. 

A direct implication of the sustained coherence is the reversibility of the
amplification. If $\Delta \varepsilon $ is small, the second term in eq.\ 
\ref{4wmeq} becomes negligible and the overall evolution is solely
determined by the unitary four-wave mixing term. By changing the sign of one
of the four waves, we can apply the inverse operator and the amplified
populations will subsequently deamplify to zero due to the entanglement
between the coupled modes. This is true even for the modes which were not
deliberately seeded, since spontaneous emission into these modes also occurs
by the same pairwise collision process. A similar experiment using photons
has already been performed \cite{Lama2001}.

After 80 $\mu $s of amplification, we initiated deamplification by inducing
a 180$^{\circ }$ phase shift in \textbf{k$_{2}$}. The sign reversal was
achieved by applying a 120 $\mu $s $\pi $-pulse resonant with the momentum
difference between the source waves \textbf{k$_{1}$} and \textbf{k$_{2}$}.
(For this experiment, also the $\pi /2$-pulse which originally generated 
\textbf{k}$_{2}$ was prolonged to 60 $\mu $s.) To check that the
amplification process was not permanently disrupted by the Bragg pulse, we
also performed a control experiment where the $\pi $-pulse was replaced by a 
$\pi /2$ and $-\pi /2$ pulse applied in quick succession. Like the first
half of the $\pi $-pulse, the $\pi /2$ pulse transferred all the atoms from 
\textbf{k$_{2}$} back into \textbf{k$_{1}$}, but unlike the second half of
the $\pi $-pulse, the $-\pi /2$ pulse built up mode occupation in \textbf{k$%
_{2}$} with the original phase. If the Bragg pulses did not otherwise
disturb the phase relation of the system, amplification should then resume.
In all three cases, we monitored the evolution of \textbf{k$_{4}$} by
applying the detection pulse and counting the number of atoms in \textbf{k$%
_{4}^{\prime }$} after variable intervals.

\begin{figure}[tbp]
\includegraphics[bb=0.5in 3.5in 7.5in 9.5in,width=60mm]{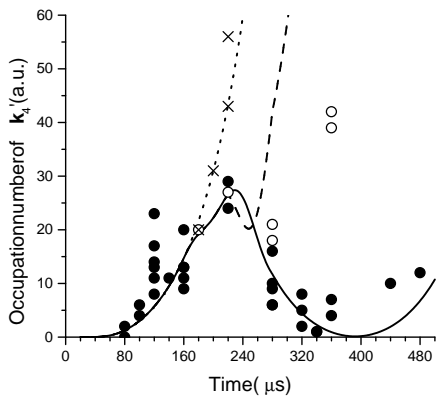}
\caption{Phase coherent amplification and deamplification: ($\bullet $)
Deamplification of the conjugate wave. ($\circ $) Continued amplification
due to a $\protect\pi $-pulse with a 180$^{\circ }$ phase shift applied in
mid-pulse (see text). ($\times $) Uninterrupted amplification. The solid,
dashed and dotted lines are independent numerical simulations of the
respective processes.}
\label{deampfig}
\end{figure}

Fig.~\ref{deampfig} shows the results obtained and clearly demonstrates the
reversal of the amplification process. As a quantitive measure of the
deamplification process, we defined the deamplification factor as the ratio
of the maximum mode occupation to the remnant left at the end of
deamplification, and obtained a factor of $\sim$~5. In addition, we infer
that this reversal is phase-dependent, since amplification continues if we
do not reverse the phase of the second source wave (Fig.~\ref{deampfig}($%
\circ $, $\times $)). This demonstrates that the four-wave mixing process
does not suffer significant decoherence during the observation time.

In Fig.~\ref{deampfig}, we also plot numerical simulations of all three
cases and find good qualitative agreement with the experiment. For these
simulations, the energy mismatch $\Delta \varepsilon $ was the main
adjustable parameter and $a_{1}$ and $a_{2}$ were taken to be c-numbers
since they are highly occupied relative to the other modes. Apart from $%
\Delta \varepsilon $, the only other adjustable parameters in the
simulations were an overall normalization constant and the size of the
initial seed, which was assumed to be small. Numerically, optimal
deamplification was achieved for $\Delta \varepsilon $ = $h$ $\times $ 1.0
kHz, and adjustments to this value by $h$ $\times $ 0.6 kHz either way still
yielded a fit consistent with the experimental data. The simulations also
took into account the off-resonant scattering of the seed and conjugate wave
during the $\pi $-pulse, which caused a 20\% loss of atoms from the
amplified waves and accounted for the dip in mode occupation seen in Fig.~3($%
\circ $). Finally, beyond 300 $\mu $s, amplification slows down in Fig.~\ref
{deampfig}($\circ $) and departs from the numerical simulation (dashed line)
due to depletion of the source waves as a result of scattering and
subsequent amplification of the unseeded modes \cite{Voge2002}.

Theoretically, maximum deamplification should occur for $\Delta \varepsilon $
= 0 in the absence of any extraneous phase shifts. Our experiment deviates
from this ideal situation as it suffered from off-resonant scattering during
the $\pi $-pulse resulting in a calculated phase shift of 0.6 rad in the
amplified modes. This was due to the finite duration of the $\pi $-pulse,
which off-resonantly coupled out atoms from the amplified waves. The
observed value of $\Delta \varepsilon $ = $h$ $\times $ 1.0 kHz therefore
might have compensated for this phase shift as the amplified modes evolved
over time.

At the same time, we also discovered that the deamplification process is
extremely phase-sensitive. Compared to the tolerance in $\Delta \varepsilon $
for amplification \cite{Voge2002}, the $h$ $\times $ 0.6 kHz tolerance for
deamplification is relatively small. While amplification is inherently a
stable process which enforces a certain phase relationship between the four
waves, deamplification is unstable with respect to the phase. As soon as the
phase relation is lost, deamplification ends and amplification resumes.
Fortunately, deamplification remains possible since its tolerance is still
larger than the estimated $h$ $\times $ 0.4 kHz halfwidth of the
inhomogeneous mean field shift. However, in order to remain within the
narrow range of $\Delta \varepsilon $ allowed, we had to align the Bragg
beams to within milliradians of the optimal value.

\begin{figure}[tbp]
\includegraphics[bb=0.5in 3.0in 7.5in 9.5in,width=60mm]{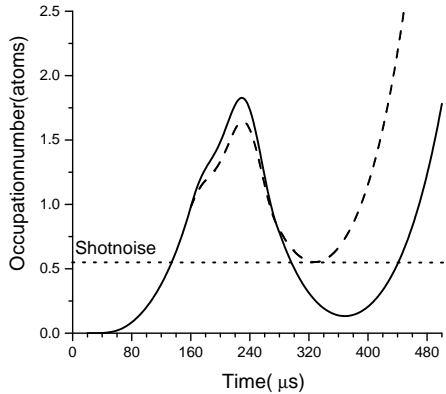}
\caption{Deamplification of perfectly squeezed modes with off-resonant
scattering (solid line). Identical process but with coherent waves (dashed
line). The vertical scale is the expectation value of the mode occupation.}
\label{cohfig}
\end{figure}

In principle, the pairwise collisions of atoms during deamplification could
be used to demonstrate relative number squeezing and entanglement between
the two amplified modes. If decoherence is absent, the mode occupations
after deamplification will reflect the relative number uncertainty in the
initial waves. For coherent or classical waves, the remnant will follow
Poissonian statistics and be of order $\sqrt{N}$, where $N$ is the initial
mode occupation. In contrast, for perfect two mode number squeezing, the
remnant will always be exactly zero. Moreover, complete deamplification of
the seed and conjugate waves can only occur if the sum of their individual
phases remain well-defined throughout the process. These two conditions
cannot be met for unentangled waves, since their individual number and phase
are canonically conjugate variables and cannot simultaneously be
well-defined.

To illustrate this, we simulated the evolution of the conjugate wave for our
experimental conditions, this time without an initial seed (Fig.~\ref{cohfig}%
a). Here, complete deamplification does not occur as the squeezing has been
degraded by off-resonant scattering. However, it still dips below the shot
noise level given by the minimum of the control graph (Fig.~\ref{cohfig}b)
which assumes coherent waves. The control simulation is started at 160 $\mu $%
s, immediately preceding the $\pi $-pulse, using non-entangled coherent
waves with the same unity mode occupation. To detect relative number
squeezing, we need to obtain deamplification factors of greater than $\sqrt{N%
}$.

Experimentally, direct detection of relative number squeezing in our
strongly occupied beams is not possible since it would require a much larger
deamplification factor than what we currently observe. In this, we are
limited by the inhomogeneous mean field and the depletion of the source
waves. However, for initially unseeded modes, a deamplification factor of
five is already evidence for sub-shot noise correlations (Fig.~\ref{cohfig}%
). This shows that by monitoring a collection of modes with a similar energy
mismatch, detection of the entanglement is possible with current
experimental parameters. Practically, however, the occupation in these modes
was masked by surrounding modes with a different energy mismatch. Further
complicating detection, the momenta of the states were blurred out during
the motion of the outcoupled atoms in the gradient of the mean field of the
condensate.

Finally, we briefly consider other possible sources of decoherence in our
amplifier and discuss how their effects may be minimized. Collisions between
the source and amplified waves can compromise squeezing as atoms are
scattered into or out of the amplified waves in a random way. As shown in 
\cite{Voge2002}, at low density, no stimulated scattering occurs and the
spontaneous scattering rate was estimated to be significant only after the
source waves are about to be depleted. By further manipulating the size of
the condensate such that the amplified waves separate out before the source
waves are depleted, we can ensure that the squeezing will be preserved.

A further study of the non-degenerate matter wave amplifier should explore
the issue of the ``beam quality'' of the squeezed waves. The phase errors
incurred due to the inhomogeneous mean field induces spatially dependent
phase shifts in the wavefunction of the squeezed waves, rendering them hard
to isolate. Other sources of phase errors are not a concern as they appear
to have little effect on our experiment. In particular, phase fluctuations
across the condensate (first observed by \cite{Dett2001}) are absent for our
parameters \cite{andr97int,sten99brag}. Moreover, since the atoms travel at
a maximum velocity of 20 mm/s for a duration of 300 $\mu $s, the required
coherence length of 6 $\mu $m is easily satisfied.

The collisional matter wave amplifier studied here complements the amplifier
based on optical superradiance \cite{inou99mwa, kozu99amp}. While the gain
and interference contrast of both are comparable, the collisional amplifier
has a much larger momentum bandwidth, amounting to a significant fraction of
the speed of sound \cite{Voge2002}. In contrast, the optical amplifier has a
bandwidth of only a single mode \cite{Moor99super}.

In conclusion, we have demonstrated the coherence of a non-degenerate matter
wave amplifier by measuring the phase of the conjugate wave and by reversing
the amplification process. While we do not yet directly detect
number-squeezing in our amplified beams, we have evaluated all known
decoherence processes and concluded that they are well controlled in our
experiment. Therefore the atomic non-degenerate parametric amplifier studied
here is a promising candidate for generating entangled and number squeezed
beams.

We thank K.\ Xu and J.\ Abo-Shaeer for experimental assistance. We would
also like to acknowledge Z.\ Hadzibabic, A.\ Chikkatur and K.\ Dieckmann for
their valuable comments and insightful discussions. This work was funded by
ONR, NSF, ARO, NASA, and the David and Lucile Packard Foundation.


\begin{thebibliography}{16}
\expandafter\ifx\csname natexlab\endcsname\relax\def\natexlab#1{#1}\fi
\expandafter\ifx\csname bibnamefont\endcsname\relax
  \def\bibnamefont#1{#1}\fi
\expandafter\ifx\csname bibfnamefont\endcsname\relax
  \def\bibfnamefont#1{#1}\fi
\expandafter\ifx\csname citenamefont\endcsname\relax
  \def\citenamefont#1{#1}\fi
\expandafter\ifx\csname url\endcsname\relax
  \def\url#1{\texttt{#1}}\fi
\expandafter\ifx\csname urlprefix\endcsname\relax\def\urlprefix{URL }\fi
\providecommand{\bibinfo}[2]{#2}
\providecommand{\eprint}[2][]{\url{#2}}

\bibitem[{\citenamefont{Deng et~al.}(1999)\citenamefont{Deng, Hagley, Wen,
  Trippenbach, Band, Julienne, Simsarian, Helmerson, Rolston, and
  Phillips}}]{deng99}
\bibinfo{author}{\bibfnamefont{L.}~\bibnamefont{Deng}},
  \bibinfo{author}{\bibfnamefont{E.~W.} \bibnamefont{Hagley}},
  \bibinfo{author}{\bibfnamefont{J.}~\bibnamefont{Wen}},
  \bibinfo{author}{\bibfnamefont{M.}~\bibnamefont{Trippenbach}},
  \bibinfo{author}{\bibfnamefont{Y.}~\bibnamefont{Band}},
  \bibinfo{author}{\bibfnamefont{P.~S.} \bibnamefont{Julienne}},
  \bibinfo{author}{\bibfnamefont{J.~E.} \bibnamefont{Simsarian}},
  \bibinfo{author}{\bibfnamefont{K.}~\bibnamefont{Helmerson}},
  \bibinfo{author}{\bibfnamefont{S.~L.} \bibnamefont{Rolston}},
  \bibnamefont{and} \bibinfo{author}{\bibfnamefont{W.~D.}
  \bibnamefont{Phillips}}, \bibinfo{journal}{Nature}
  \textbf{\bibinfo{volume}{398}}, \bibinfo{pages}{218} (\bibinfo{year}{1999}).

\bibitem[{\citenamefont{Vogels et~al.}(2002)\citenamefont{Vogels, Xu, and
  Ketterle}}]{Voge2002}
\bibinfo{author}{\bibfnamefont{J.~M.} \bibnamefont{Vogels}},
  \bibinfo{author}{\bibfnamefont{K.}~\bibnamefont{Xu}}, \bibnamefont{and}
  \bibinfo{author}{\bibfnamefont{W.}~\bibnamefont{Ketterle}},
  \bibinfo{journal}{Phys. Rev. Lett.} \textbf{\bibinfo{volume}{89}},
  \bibinfo{pages}{020401} (\bibinfo{year}{2002}).

\bibitem[{not()}]{note1}
\bibinfo{note}{The atomic amplifier is non-degenerate in the sense that both
  amplified modes are distinct. This should not be confused with the energetic
  degeneracy of all four waves in the center-of-mass frame.}

\bibitem[{\citenamefont{Holland and Burnett}(1993)}]{holl93}
\bibinfo{author}{\bibfnamefont{M.~J.} \bibnamefont{Holland}} \bibnamefont{and}
  \bibinfo{author}{\bibfnamefont{K.}~\bibnamefont{Burnett}},
  \bibinfo{journal}{Phys. Rev. Lett.} \textbf{\bibinfo{volume}{71}},
  \bibinfo{pages}{1355} (\bibinfo{year}{1993}).

\bibitem[{\citenamefont{Bouyer and Kasevich}(1997)}]{bouy97}
\bibinfo{author}{\bibfnamefont{P.}~\bibnamefont{Bouyer}} \bibnamefont{and}
  \bibinfo{author}{\bibfnamefont{M.~A.} \bibnamefont{Kasevich}},
  \bibinfo{journal}{Phys. Rev. A} \textbf{\bibinfo{volume}{56}},
  \bibinfo{pages}{R1083} (\bibinfo{year}{1997}).

\bibitem[{\citenamefont{Greiner et~al.}()\citenamefont{Greiner, Mandel,
  Haensch, and Bloch}}]{Grein2002}
\bibinfo{author}{\bibfnamefont{M.}~\bibnamefont{Greiner}},
  \bibinfo{author}{\bibfnamefont{O.}~\bibnamefont{Mandel}},
  \bibinfo{author}{\bibfnamefont{T.~W.} \bibnamefont{Haensch}},
  \bibnamefont{and} \bibinfo{author}{\bibfnamefont{I.}~\bibnamefont{Bloch}},
  \bibinfo{note}{preprint cond/mat 0207196}.

\bibitem[{\citenamefont{Dalfovo et~al.}(1999)\citenamefont{Dalfovo, Giorgini,
  Pitaevskii, and Stringari}}]{Dalf1999}
\bibinfo{author}{\bibfnamefont{F.}~\bibnamefont{Dalfovo}},
  \bibinfo{author}{\bibfnamefont{S.}~\bibnamefont{Giorgini}},
  \bibinfo{author}{\bibfnamefont{L.~P.} \bibnamefont{Pitaevskii}},
  \bibnamefont{and}
  \bibinfo{author}{\bibfnamefont{S.}~\bibnamefont{Stringari}},
  \bibinfo{journal}{Rev. Mod. Phys.} \textbf{\bibinfo{volume}{71}},
  \bibinfo{pages}{463} (\bibinfo{year}{1999}).

\bibitem[{\citenamefont{Dettmer et~al.}(2001)\citenamefont{Dettmer, Hellweg,
  Ryytty, Arlt, Ertmer, Sengstock, Petrov, Shlyapnikov, Kreutzmann, Santos
  et~al.}}]{Dett2001}
\bibinfo{author}{\bibfnamefont{S.}~\bibnamefont{Dettmer}},
  \bibinfo{author}{\bibfnamefont{D.}~\bibnamefont{Hellweg}},
  \bibinfo{author}{\bibfnamefont{P.}~\bibnamefont{Ryytty}},
  \bibinfo{author}{\bibfnamefont{J.~J.} \bibnamefont{Arlt}},
  \bibinfo{author}{\bibfnamefont{W.}~\bibnamefont{Ertmer}},
  \bibinfo{author}{\bibfnamefont{K.}~\bibnamefont{Sengstock}},
  \bibinfo{author}{\bibfnamefont{D.~S.} \bibnamefont{Petrov}},
  \bibinfo{author}{\bibfnamefont{G.~V.} \bibnamefont{Shlyapnikov}},
  \bibinfo{author}{\bibfnamefont{H.}~\bibnamefont{Kreutzmann}},
  \bibinfo{author}{\bibfnamefont{L.}~\bibnamefont{Santos}},
  \bibnamefont{et~al.}, \bibinfo{journal}{Phys. Rev. Lett.}
  \textbf{\bibinfo{volume}{87}}, \bibinfo{pages}{160406}
  (\bibinfo{year}{2001}).

\bibitem[{\citenamefont{Kozuma et~al.}(1999{\natexlab{a}})\citenamefont{Kozuma,
  Deng, Hagley, Wen, Lutwak, Helmerson, Rolston, and Phillips}}]{kozu1999}
\bibinfo{author}{\bibfnamefont{M.}~\bibnamefont{Kozuma}},
  \bibinfo{author}{\bibfnamefont{L.}~\bibnamefont{Deng}},
  \bibinfo{author}{\bibfnamefont{E.~W.} \bibnamefont{Hagley}},
  \bibinfo{author}{\bibfnamefont{J.}~\bibnamefont{Wen}},
  \bibinfo{author}{\bibfnamefont{R.}~\bibnamefont{Lutwak}},
  \bibinfo{author}{\bibfnamefont{K.}~\bibnamefont{Helmerson}},
  \bibinfo{author}{\bibfnamefont{S.~L.} \bibnamefont{Rolston}},
  \bibnamefont{and} \bibinfo{author}{\bibfnamefont{W.~D.}
  \bibnamefont{Phillips}}, \bibinfo{journal}{Phys. Rev. Lett.}
  \textbf{\bibinfo{volume}{82}}, \bibinfo{pages}{871}
  (\bibinfo{year}{1999}{\natexlab{a}}).

\bibitem[{\citenamefont{Chikkatur et~al.}(2000)\citenamefont{Chikkatur,
  Görlitz, Stamper-Kurn, Inouye, Gupta, and Ketterle}}]{chik00}
\bibinfo{author}{\bibfnamefont{A.~P.} \bibnamefont{Chikkatur}},
  \bibinfo{author}{\bibfnamefont{A.}~\bibnamefont{Görlitz}},
  \bibinfo{author}{\bibfnamefont{D.~M.} \bibnamefont{Stamper-Kurn}},
  \bibinfo{author}{\bibfnamefont{S.}~\bibnamefont{Inouye}},
  \bibinfo{author}{\bibfnamefont{S.}~\bibnamefont{Gupta}}, \bibnamefont{and}
  \bibinfo{author}{\bibfnamefont{W.}~\bibnamefont{Ketterle}},
  \bibinfo{journal}{Phys. Rev. Lett.} \textbf{\bibinfo{volume}{85}},
  \bibinfo{pages}{483} (\bibinfo{year}{2000}).

\bibitem[{\citenamefont{Lamas-Linares et~al.}(2001)\citenamefont{Lamas-Linares,
  Howell, and Bouwmeester}}]{Lama2001}
\bibinfo{author}{\bibfnamefont{A.}~\bibnamefont{Lamas-Linares}},
  \bibinfo{author}{\bibfnamefont{J.~C.} \bibnamefont{Howell}},
  \bibnamefont{and}
  \bibinfo{author}{\bibfnamefont{D.}~\bibnamefont{Bouwmeester}},
  \bibinfo{journal}{Nature} \textbf{\bibinfo{volume}{412}},
  \bibinfo{pages}{887} (\bibinfo{year}{2001}).

\bibitem[{\citenamefont{Andrews et~al.}(1997)\citenamefont{Andrews, Townsend,
  Miesner, Durfee, Kurn, and Ketterle}}]{andr97int}
\bibinfo{author}{\bibfnamefont{M.~R.} \bibnamefont{Andrews}},
  \bibinfo{author}{\bibfnamefont{C.~G.} \bibnamefont{Townsend}},
  \bibinfo{author}{\bibfnamefont{H.-J.} \bibnamefont{Miesner}},
  \bibinfo{author}{\bibfnamefont{D.~S.} \bibnamefont{Durfee}},
  \bibinfo{author}{\bibfnamefont{D.~M.} \bibnamefont{Kurn}}, \bibnamefont{and}
  \bibinfo{author}{\bibfnamefont{W.}~\bibnamefont{Ketterle}},
  \bibinfo{journal}{Science} \textbf{\bibinfo{volume}{275}},
  \bibinfo{pages}{637} (\bibinfo{year}{1997}).

\bibitem[{\citenamefont{Stenger et~al.}(1999)\citenamefont{Stenger, Inouye,
  Chikkatur, Stamper-Kurn, Pritchard, and Ketterle}}]{sten99brag}
\bibinfo{author}{\bibfnamefont{J.}~\bibnamefont{Stenger}},
  \bibinfo{author}{\bibfnamefont{S.}~\bibnamefont{Inouye}},
  \bibinfo{author}{\bibfnamefont{A.~P.} \bibnamefont{Chikkatur}},
  \bibinfo{author}{\bibfnamefont{D.~M.} \bibnamefont{Stamper-Kurn}},
  \bibinfo{author}{\bibfnamefont{D.~E.} \bibnamefont{Pritchard}},
  \bibnamefont{and} \bibinfo{author}{\bibfnamefont{W.}~\bibnamefont{Ketterle}},
  \bibinfo{journal}{Phys. Rev. Lett.} \textbf{\bibinfo{volume}{82}},
  \bibinfo{pages}{4569} (\bibinfo{year}{1999}).

\bibitem[{\citenamefont{Inouye et~al.}(1999)\citenamefont{Inouye, Pfau, Gupta,
  Chikkatur, Görlitz, Pritchard, and Ketterle}}]{inou99mwa}
\bibinfo{author}{\bibfnamefont{S.}~\bibnamefont{Inouye}},
  \bibinfo{author}{\bibfnamefont{T.}~\bibnamefont{Pfau}},
  \bibinfo{author}{\bibfnamefont{S.}~\bibnamefont{Gupta}},
  \bibinfo{author}{\bibfnamefont{A.~P.} \bibnamefont{Chikkatur}},
  \bibinfo{author}{\bibfnamefont{A.}~\bibnamefont{Görlitz}},
  \bibinfo{author}{\bibfnamefont{D.~E.} \bibnamefont{Pritchard}},
  \bibnamefont{and} \bibinfo{author}{\bibfnamefont{W.}~\bibnamefont{Ketterle}},
  \bibinfo{journal}{Nature} \textbf{\bibinfo{volume}{402}},
  \bibinfo{pages}{641} (\bibinfo{year}{1999}).

\bibitem[{\citenamefont{Kozuma et~al.}(1999{\natexlab{b}})\citenamefont{Kozuma,
  Suzuki, Torii, Sugiura, Kuga, Hagley, and Deng}}]{kozu99amp}
\bibinfo{author}{\bibfnamefont{M.}~\bibnamefont{Kozuma}},
  \bibinfo{author}{\bibfnamefont{Y.}~\bibnamefont{Suzuki}},
  \bibinfo{author}{\bibfnamefont{Y.}~\bibnamefont{Torii}},
  \bibinfo{author}{\bibfnamefont{T.}~\bibnamefont{Sugiura}},
  \bibinfo{author}{\bibfnamefont{T.}~\bibnamefont{Kuga}},
  \bibinfo{author}{\bibfnamefont{E.~W.} \bibnamefont{Hagley}},
  \bibnamefont{and} \bibinfo{author}{\bibfnamefont{L.}~\bibnamefont{Deng}},
  \bibinfo{journal}{Science} \textbf{\bibinfo{volume}{286}},
  \bibinfo{pages}{2309} (\bibinfo{year}{1999}{\natexlab{b}}).

\bibitem[{\citenamefont{Moore and Meystre}(1999)}]{Moor99super}
\bibinfo{author}{\bibfnamefont{M.~G.} \bibnamefont{Moore}} \bibnamefont{and}
  \bibinfo{author}{\bibfnamefont{P.}~\bibnamefont{Meystre}},
  \bibinfo{journal}{Phys. Rev. Lett.} \textbf{\bibinfo{volume}{83}},
  \bibinfo{pages}{5202} (\bibinfo{year}{1999}).

\end{thebibliography}
\end{document}